\documentstyle[11pt,paspconf,epsf]{article}

\def\etal{et al.\ }
\def\vmi{\hbox{\it V--I\/}}
\def\bu{$\bullet$}

\def\kms{km s$^{-1}$}                       
\def\kmsm{km s$^{-1}$ Mpc$^{-1}$}

\def\h0{$H_0$}

\begin{document}

\title{The Hubble Constant from the HST Key Project on the Extragalactic Distance Scale}

\author{Laura Ferrarese\altaffilmark{1},
Brad K. Gibson\altaffilmark{2}, 
Daniel D. Kelson\altaffilmark{3},
Shoko Sakai\altaffilmark{4},
Jeremy R. Mould\altaffilmark{5}, 
Wendy L. Freedman\altaffilmark{6}, 
Robert C. Kennicutt, Jr.\altaffilmark{7}, 
Holland C. Ford\altaffilmark{8}, 
John A. Graham\altaffilmark{3}, 
John Huchra\altaffilmark{9}, 
Shaun M. Hughes\altaffilmark{10}, 
Garth D. Illingworth\altaffilmark{11},  
Lucas Macri\altaffilmark{9}, 
Barry F. Madore\altaffilmark{12}, 
Kim Sebo\altaffilmark{5}, 
N.A. Silbermann\altaffilmark{12}  
\& Peter B. Stetson\altaffilmark{13}}

\altaffiltext{1}{Hubble Fellow, California Institute of Technology, Pasadena CA 91125, USA; 
$^{2}${CASA, University of Colorado, Boulder, CO, USA};
$^{3}${Department of Terrestrial Magnetism, Carnegie Institution of
Washington, Washington DC 20015, USA};
$^{4}${Kitt Peak National Observatory, NOAO, Tucson AZ 85726, USA};
$^{5}${Research School of Astronomy \& Astrophysics, Institute of
Advanced Studies, ANU, ACT 2611, Australia};
$^{6}${Carnegie Observatories, Pasadena CA 91101, USA};
$^{7}${Steward Observatory, The University of Arizona, Tucson AZ 85721,
USA};
$^{8}${Johns Hopkins University and Space Telescope
Science Institute, Baltimore MD 21218, USA};
$^{9}${Harvard Smithsonian Center for Astrophysics, Cambridge MA 02138
USA};
$^{10}${Royal Greenwich Observatory, Cambridge CB3 OHA, UK};
$^{11}${Lick Observatory, University of California, Santa Cruz CA 95064
USA};
$^{12}${NASA/IPAC Extragalactic Database and California Institute of
Technology, Pasadena CA 91125, USA};
$^{13}${Dominion Astrophysical Observatory, Victoria, British Columbia
V8X 4M6, Canada}}

\begin{abstract}
The final efforts of the HST Key Project on the Extragalactic Distance
Scale are presented. Four distance indicators, the Surface Brightness
Fluctuation method, the Fundamental Plane for early-type galaxies, the
Tully-Fisher relation and the Type Ia Supernovae, are calibrated
using Cepheid distances to galaxies within 25 Mpc. The calibration is
then applied to distant samples reaching $cz \sim 10000$ \kms~ and (in the case
of SNIa) beyond. By combining the constraints imposed on the Hubble
constant by the four distance indicators, we obtain \h0 = 71$\pm$6
\kmsm.
\end{abstract}

\section{Introduction}

Back in 1984, the goal of the HST Key Project on the Extragalactic
Distance Scale (hereafter, KP) was announced to be the measurement of
the Hubble constant, \h0, with 10\% accuracy. The plan of attack was
to set the zero points for a variety of secondary distance indicators
by measuring distances to 18 nearby calibrators using the most
reliable of the primary standard candles,  Cepheid variables (Figure
1). Crucial to the success of the mission was the ability of secondary
distance indicators to reach beyond the local supercluster into the
unperturbed Hubble flow.  Fifteen years later, four have proven up to
the challenge: the Surface Brightness Fluctuation Method, the
Fundamental Plane for early-type galaxies, the Tully-Fisher
relation and the Type Ia Supernovae. All are subject to biases arising
from implicit assumptions made on the stellar population of the
galaxies they target.  Such biases, which could have serious effects
if any one distance indicator were to be used alone,  are attenuated
when the constraints imposed by all indicators are combined.  The
final KP results are presented in Ferrarese \etal (2000a), Kelson \etal (2000), 
Sakai \etal (2000), Gibson \etal (2000), Mould \etal
(2000) and Freedman \etal (2000).  Here we will summarize those
efforts, focusing on the merit and disadvantages of each indicators,
and pointing out areas where future work is needed.

The following sets the scene for the remainder of this paper: the
calibration of the PL relation is based on the LMC sample of Madore \&
Freedman (1991).  It assumes a true distance modulus to the LMC of
$18.50 \pm 0.13$ mag (Mould \etal 2000),  no dependence of the
Cepheid PL relation on the metallicity of the variable stars, a ratio
of total to selective absorption $R_V=A(V)/E(B-V)=3.3$, and a
reddening law following Cardelli, Clayton and Mathis (1989). Each of
these assumptions is examined and their effects on the final error
budget are assessed in the last section. We strove for homogeneity
between the Cepheid distance scale and all secondary distance
indicators: the treatment of the calibrator sample is consistent
internally and with the distant sample to which the calibration is
applied (Ferrarese \etal 2000b, Macri \etal 2000, Gibson \etal
2000, Kelson \etal 2000), providing a meaningful comparison of
the results. This is perhaps the most distinctive mark of the KP
compared to previous work.

\begin{table}
\caption{Cepheid Calibrators for the Secondary Distance Scale} \label{tbl-1}
\begin{center}\scriptsize
\begin{tabular}{lllccccl}
Galaxy & m-M (mag)\tablenotemark{a} & Group/Cluster\tablenotemark{b} & SBF\tablenotemark{c} & FP\tablenotemark{c} & TF\tablenotemark{c} & SNIa\tablenotemark{c} & Reference\tablenotemark{d}\\ 
\tableline
N224     & 24.44$\pm$0.10 & Local Group    &\bu&   &\bu&   &Madore \& Freedman 1991 \\  
N598     & 24.64$\pm$0.09 & Local Group    &   &   &\bu&   & Freedman \etal 1991 \\ 
N925     & 29.84$\pm$0.08 & N1023 Group   &   &   &\bu &   & Silbermann \etal 1996 \\  
N1326A   & 31.43$\pm$0.07 & Fornax Cluster&   &\bu&   &   & Prosser \etal 1999 \\  
N1365 	 & 31.39$\pm$0.10 & Fornax Cluster&   &\bu&\bu&   & Silbermann \etal 1999 \\
N1425 	 & 31.81$\pm$0.06 & Fornax Cluster&   &\bu&\bu&   & Mould \etal 1999 \\
N2090 	 & 30.45$\pm$0.08 & N1808 Group   &   &   &\bu&   & Phelps \etal 1998 \\
N2541 	 & 30.47$\pm$0.08 & N2841 Group   &   &   &\bu&   & Ferrarese \etal 1998 \\
N2403    & 27.51$\pm$0.24 & M81 region    &   &   &\bu&   & Madore \& Freedman 1991 \\
N3031 	 & 27.80$\pm$0.08 & M81 Group     &\bu&   &\bu&   & Freedman \etal 1994 \\
N3198 	 & 30.80$\pm$0.06 & N3184 Group        &   &   &\bu&   & Kelson \etal 1999 \\
N3319 	 & 30.78$\pm$0.10 & N3184 Group   &   &   &\bu&   & Sakai \etal 1999 \\
N3351 	 & 30.01$\pm$0.08 &Leo I Group    &   &\bu&\bu&   & Graham \etal 1997 \\
N3368 	 &30.20$\pm$0.10 & Leo I Group    &\bu&\bu&\bu&\bu& Gibson \etal  2000\\
N3627 	 & 30.06$\pm$0.17 & Leo region    &   &   &\bu&\bu& Gibson \etal  2000\\
N3621 	 & 29.13$\pm$0.11 & Isolated      &   &   &\bu&   & Rawson \etal 1997 \\
N4414 	 &31.41$\pm$0.10 & Coma Clouds    &   &   &\bu&$\circ$& Turner \etal 1998 \\
N4725 	 & 30.57$\pm$0.08 & Coma Clouds   &\bu&   &\bu&   & Gibson \etal 1998 \\
IC4182 	 & 28.36$\pm$0.08 & Coma Clouds region&   &   &   &\bu& Gibson \etal 2000 \\
N4321 	 & 31.04$\pm$0.09 & Virgo$-$M87   &   &\bu&   &   & Ferrarese  \etal 1996 \\
N4548 	 & 31.04$\pm$0.08 &Virgo$-$M87    &\bu&\bu&\bu&   & Graham \etal 1998 \\
N4496A   & 31.02$\pm$0.07 & Virgo$-$N4472 &   &\bu&   &$\circ$& Gibson \etal 2000\\
N4535 	 & 31.10$\pm$0.07 & Virgo$-$N4472 &   &\bu&\bu&   & Macri \etal 1999 \\
N4536 	 & 30.95$\pm$0.07 &Virgo$-$N4472  &   &\bu&\bu&\bu& Gibson \etal 2000 \\
N4639 	 & 31.80$\pm$0.09 & Virgo$-$N4649 &   &$\circ$&   &\bu& Gibson \etal 2000 \\
N5253 	 & 27.61$\pm$0.11 & Cen A Group    &   &   &   &\bu& Gibson \etal 2000 \\
N7331 	 & 30.89$\pm$0.10 & N7331 Group   &\bu&   &\bu&   & Hughes \etal 1998 \\
\tableline
\end{tabular}
\end{center}\scriptsize
\noindent $^a$Cepheid distance modulus (random errors only; the systematic error is 0.16 mag.).

\noindent $^b$Group/cluster definition is given in Ferrarese \etal
2000b. 

\noindent $^c$A bullet identifies galaxies used as calibrators for the
method. An open circle defines potential calibrators which are however
not used.
 
\noindent $^d$Full references are given in Ferrarese \etal 2000b.

\end{table}

\section{Surface Brightness Fluctuations}

The KP calibration of the Surface Brightness Fluctuation method (Tonry
\& Schneider 1988) is discussed in Ferrarese \etal (2000a).  The
largest database of SBF measurements comprises $\sim$300 galaxies
within the local supercluster,  observed from the ground in the
Kron-Cousins $I$-band by John Tonry and collaborators (Tonry \etal
1999, Ajhar \etal 2000). A much smaller ($\sim 20$ galaxies), but more
distant survey has employed the WFPC2 on board HST (Ajhar \etal 1997,
Thomsen \etal 1997, Pahre \etal 1999, Lauer \etal 1998); indeed, it is
from this pool of $\sim 20$ galaxies that the six at $cz \sim
3000-7000$ \kms~ have been singled out by the KP for deriving \h0. The
use of HST for SBF measurements has some drawbacks, however: the
calibration of the fluctuation magnitudes cannot proceed directly
against the Cepheids, since only one galaxy, M31, is in common between
the two methods. Furthermore, fluctuation magnitudes are known to
depend strongly on the metallicity (traditionally expressed as a
\vmi~color) of the underlying stellar population (Tonry \etal
1997). Such a dependence cannot be properly quantified for the HST-SBF
sample because of its limited size (Ajhar \etal 1997). Both problems
can be overcome, but they certainly are important enough to deserve
further study. As for the color dependence, the KP approach is to
assume it to be the same as determined empirically for the larger
ground based $I$-band survey (Tonry \etal 1997). This choice is
supported by stellar population synthesis models (Worthey 1994), and
the similar response curves of the $I$ and the filter used for the
HST/WFPC2 measurements, F814W. 

Once the color dependence is accounted for, the absolute magnitude of
the fluctuations measured with HST can be derived using the 16
galaxies which also have ground based $I$-band SBF measurements,
calibrated against the Cepheids. The latter calibration, however,
poses some further problems: all galaxies with $I$-SBF and Cepheid
distances (see Table 1)  are early-type spirals, for which SBF
measurements become challenging  (Tonry \etal 1999). Dust and other
contaminants can conspire to artificially brighten the measured
fluctuations, and the stellar  population in bulges might not be
identical to that of the  ellipticals which are the preferred targets
of the method (leading to  a different color dependency). On the other
hand, using only SBF data to early-type galaxies would force an
indirect calibration whose validity relies on the precarious
assumption of a spatial coincidence between late and early-type
galaxies, further aggravated by the small sample of galaxies with
Cepheid distances in any group.  Given the current data (Ferrarese 
\etal 2000b), the direct calibration for the six galaxies in
Table 1, and the indirect one for the six groups/clusters with mean
Cepheid and SBF distances (including Leo I, Virgo and Fornax), differ
by 0.1 mag, the indirect calibration leading to larger distances. To
avoid uncertainties introduced by cluster depth effects, the KP has
preferred the direct calibration, but we  stress that  the reliability
of the SBF measurements in spirals remains to be tested, and coupled
with the not fully satisfactory constraints imposed on the color
dependence of HST-SBF, is the main source of concern in the present
calibration of the SBF method.

Because the 5 galaxies comprising the SBF distant sample are  confined
within 5000 \kms (excluding the not very well constrained measurement
in the Coma cluster by Thomsen \etal 1997), SBF is more susceptible
to the effects of large scale flows than the other
indicators. Furthermore, three of the five galaxies lie in the
immediate vicinity of one of the major contributors to the local flow
field, i.e., the Great Attractor.  To provide a first order estimate of
the bias introduced by the flow field in estimating \h0, we considered
a simple multi-attractor model fully described in Mould \etal 
(2000). The model  assumes three mass concentrations, the local
supercluster, the Great Attractor, and the Shapely Concentration,
acting independently so that corrections for each are additive.  Using
velocities corrected  for this flow model, and the HST-SBF distances
derived using the direct calibration described above, we obtain \h0 =
69$\pm$4  (random) $\pm$5 (systematic) \kmsm. \h0 happens to remains
unaltered if velocities corrected to the CMB frame are used instead;
nevertheless, we estimate an 8 \kmsm~ random error on \h0 {\it per
cluster} due to corrections for the flow field.  This represents the
major contributor to the random uncertainty, random errors in the
fluctuation magnitudes and \vmi~colors carry only 30\% of the
weight. The systematic uncertainty is due mostly to the systematic
error in the Cepheid distance scale (the distance to the LMC and the
photometric calibration of the WFPC2), and partly to the internal
error in the SBF calibration.

\section{Fundamental Plane}

Among the distance indicators targeted by the KP, the Fundamental
Plane (FP) has the distinct disadvantage of not being calibratable
directly against the Cepheids. The approach followed for the KP by
Kelson \etal (2000)  is to base the FP calibration on the Leo I
group and the Fornax and Virgo clusters, for which several Cepheid distances
exist. Based on Monte Carlo simulations following Gonzales \& Faber
(1997), Kelson \etal test the hypothesis of a spatial coincidence
between the Cepheid spirals and the FP ellipticals in these clusters,
and conclude that this assumption leads to an underestimate of the FP
distances. Accurate evaluation of this bias would require  a detailed
knowledge of the clusters' 3D spatial structure which, alas, is
lacking. The simulations suggest a 5\% downward correction to \h0, but
at the very high price of a 5\% uncertainty, which alone will account
for a quarter of the systematic error budget on \h0. 

The kinematic data (i.e., the velocity dispersion) used by Kelson \etal
for the fundamental plane in Leo I, Virgo and Fornax are from
Dressler \etal (1987) and Faber \etal (1989), while the photometric
parameters (i.e., effective radius and surface brightness) are derived
from original $I$-band data. The distant sample consists of 11
clusters observed in Gunn $r$ by Jorgensen \etal (1995ab) in the range
$cz \sim 1100-12000$ \kms. For consistency, the slope of the
FP for the calibrators is assumed to be the same as for the distant
sample, the $I$ band photometry for the local calibrators is
transformed to Gunn $r$, and the velocity dispersion is corrected to the
same aperture used for the distant sample. None of these 
steps produces major contributions to the error budget.

There appear to be no additional major impasses with the  calibration:
the zero points derived from the three clusters are consistent with
each other. In view of the findings of Fruchter \etal (this volume) it
is noteworthy, even if inconsequential, that the FP dispersion is
found to be almost double in  Fornax compared to Virgo and Leo I
($\sim$0.09 dex compared to $\sim$0.05 dex). The calibration applied
to the distant sample leads to \h0=78$\pm$7 (random) $\pm$8
(systematic) \kmsm. The major source of random uncertainty is
associated with the slope and zero point of the fundamental plane  and
only in part with the Cepheid distances of the calibrating
clusters. The systematic uncertainty derives mainly from the
systematic error in the Cepheid distance scale, the uncertain
accounting of the spatial separation between spirals and ellipticals
(see above),  and only in minor part from cluster population
incompleteness bias.

\section{Tully-Fisher Relation}

The KP calibration of the $BVRIH$ Tully-Fisher relations is
presented in Sakai  \etal (2000). A summary has been written by Shoko
Sakai for these proceedings, therefore only a few words will be spent
here. The calibration, based on 21 galaxies with Cepheid distances
(Table 1), is applied to four distant cluster samples. Of these the
largest is the $I$-band survey of Giovanelli \etal (1997),  which
comprises 555 galaxies in 24 clusters within 9500 \kms. The  $B$ and
$V$-band samples of Bothun \etal (1985) and the  $H$-band sample of
Aaronson \etal (1982, 1986) span a comparable range in redshift
space, but are of significantly smaller size. The most significant
problem with the TF analysis is the discrepancy, at the 2$\sigma$
level, between the values of \h0 derived from the $I$ and $H$ surveys:
74$\pm$2  \kmsm~ and 67$\pm$3 \kmsm~ respectively (random errors
only). Sakai \etal thoroughly investigated the cause of this
discrepancy, and even though a secure culprit could not be identified,
circumstantial evidence points to the $H$-band photometry as the most
likely suspect. Work is in progress (Macri \etal 2000) to re-derive
$H$ magnitudes for the local calibrators, and resolve the $I$ {\it vs}
$H$ disagreement. At this time, the most prudent course of action is
to adopt a weighted average of the values of \h0 from all four
surveys. This leads to \h0 = 71$\pm$4 (random) $\pm$10 (systematic)
\kmsm. The random error is shared equally by errors in the CMB
velocities for the distant samples, and the random error in the
Tully-Fisher moduli (mainly deriving from errors in the photometry,
and only partly in the linewidths). The systematic uncertainties come
from the systematic error in the Cepheid distance scale and in the TF
zero point. Notice that unlike SBF and FP, which target the dust free
environments of early-type galaxies, TF and SNIa carry the extra
burden of having to deal with internal extinction, which produces an
additional term in the final error budget.

\section{Type Ia Supernovae}

A chronicle on \h0 from Type Ia Supernovae (SNIa) would be divided
into three chapters. In chapter 1, values in the upper 50s would be
routinely recorded under the  assumption that the magnitude at peak
acts as a standard  candle (Hamuy \etal 1996, Riess \etal 1996, Saha
\etal 1997). Chapter 2 opens with the realization that the intrinsic
brightness at maximum light correlates with the decline rate (as first
suggested by Phillips 1993): slow decliners are intrinsically brighter
than fast decliners. Because the local calibrators happen to be slower
decliners than the average SNIa in the distant samples, correction for
this  effect leads to a substantial 8\% increase in \h0 (Saha \etal
1999, Hamuy \etal 1996, Riess \etal 1998, Phillips \etal
1999). The KP wrote chapter 3 by re-deriving Cepheid distances
to the local SNIa host galaxies (Gibson \etal 2000), thus revising
the calibration and leading to a further 6\% increase in \h0. 

It is thanks to the effort of Allan Sandage and collaborators if we can
now rely on Cepheid distances to six nearby SNIa host galaxies when
none existed before (Saha \etal 1994, 1995, 1996a, 1996b, 1997,
1999), even if  nothing compares to the strategic planning of Tanvir
\etal (1995) who  published a Cepheid distance to NGC 3368 three
years before its SNIa went off. Loyal to the KP commitment of
providing a consistent footing to all secondary distance indicators,
Gibson \etal (2000) set out to derive new photometry and distances to
all of the SNIa host galaxies. In two cases (NGC 4496A and IC4182) the
new distances agree with the ones originally published; but in all
other cases the re-analysis led to consistently smaller distance
moduli,  by an average of 0.12 magnitudes. The causes of the
discrepancies vary from galaxy to galaxy; they include disagreement in
the photometry (which are quickly amplified by the de-reddening
procedure used in deriving the distance moduli), differences in the
sample of Cepheids, and differences in the treatment of reddening. Many
pages of justifications and details are given  in Gibson et al.; the
punch line is that these new distances should be preferred when SNIa
are compared to the other secondary distance indicators, as they are
derived in a consistent manner as for the 18 galaxies originally
observed as part of the KP.

Three local supernovae are excluded from the calibration because of
the poor quality of their light curves: SN1895B in NGC 5253 (the
galaxy also hosted the better sampled SN1972E) and the SNe in NGC 4414
and NGC 4496A.  Gibson \etal follow Suntzeff \etal (1999) in the
adoption of the SNe data ($B$, $V$ and $I$ photometry and extinction
corrections) for both the local and distant sample. The latter
comprises  35 Cal\'an-Tololo/CfA SNe (Phillips \etal 1999), in the
range $cz \sim 1000-31000$ \kms.  Both the distant and nearby samples
are corrected for the decline rate versus peak luminosity relation
obtained by Phillips \etal (1999) from a low extinction subset of the
Cal\'an-Tololo SNe. Averaging the  $B$, $V$ and $I$ data Gibson \etal
derive a weighted Hubble constant of \h0=68$\pm$2 (random) $\pm$5
(systematic) \kmsm. The main contributions to the random errors come
from the scatter in the Hubble diagram and  from errors in the
photometry and distances for the local calibrators. The systematic
error is propagated directly from the Cepheid distance scale.

How does the KP value of \h0 compare to the findings of other groups?
We will consider the two most recent works. Suntzeff \etal (1999),
which is the source of the KP database for both the local and distant
SNe,  quote \h0 = 63$\pm$2 \kmsm which, as expected, agrees with
Gibson \etal  when the new distances for the local calibrators are
accounted for. Saha \etal (1999) obtain \h0 = 60$\pm$2 \kmsm~ in both
$B$ and $V$. Part of the difference with Gibson \etal is again due to
the adoption of new distances, but an additional 6\% must be accounted 
for.  This is due to differences in the internal extinction
corrections for the local sample, and in the foreground extinction for
the distant sample (the two having opposite effects on \h0), and in
the actual photometry adopted for the local SNe. Other factors, such
as the adoption of slightly different distant samples, different
formalism for the rate of decline versus peak magnitude relation, and
different assumption as to a dependence of \h0 on redshift, do not
produce appreciable differences.

\section{Combining the Constraints}

By combining the constraints imposed on \h0 by each
indicator (Figure 2), we can reduce the propagation of  uncorrelated
systematics. There is some amount of covariance in the values of \h0
from Table 2,   due for example to the sharing of some of the
calibrator galaxies (Table 1), and to the common underlying assumption
on the distance to the LMC. To properly account for the interplay of
random and systematic errors, Mould \etal (2000) have developed Monte
Carlo simulations in which all uncertainties and parameter dependences
for each indicator are propagated through and investigated
thoroughly. {\bf The final result is \h0=71$\pm$6 \kmsm. The error
distribution is symmetrical, with a 1$\sigma$ width of 9\%, fulfilling
the KP original goal.}

\begin{table}
\caption{\h0 from Secondary Distance Indicators} \label{tbl-1}
\begin{center}\scriptsize
\begin{tabular}{lccccccccl}
 Indicator & $\sigma$\tablenotemark{a} & No.\tablenotemark{b} & No.\tablenotemark{b} & $\Delta$v\tablenotemark{c} &\h0\tablenotemark{d}  & \multicolumn{3}{c}{Systematics\tablenotemark{e}} \\
\tableline 
%
%
%
SBF & 0.11 & 6 & 4 & 3900$-$4700 & 69$\pm$4$\pm$6 & $1\%\Uparrow$ &  $5\%\Downarrow$ & N/A \\

FP  & 0.045 & 16 &11 & 1100$-$12000& 78$\pm$7$\pm$8 & $1\%\Uparrow$ &  $6\%\Downarrow$ & 2\% \\

TF& 0.19 & 21 &24 &1000$-$11500&71$\pm$4$\pm$10 & $1\%\Uparrow$ &  $4\%\Downarrow$ & 2\% \\

SNIa& 0.14 & 6 &35 &1000$-$31000&68$\pm$2$\pm$5 & $2\%\Uparrow$ &  $4\%\Downarrow$ & N/A \\
\tableline
{\bf Combined} & & & & & {\bf71$\pm$6} & $1\%\Uparrow$ &  $5\%\Downarrow$ &  & \\
\tableline
\end{tabular}
\end{center}\scriptsize
\vskip -.01in
\noindent $^a$The intrinsic dispersion for each indicator, in magnitudes for TF, SBF and SNIa, and dex for FP. 

\noindent $^b$The number of local calibrators and of clusters/galaxies in the distant sample respectively.

\noindent $^c$Velocity range (in \kms) spanned by the distant sample. 

\noindent $^d$In \kmsm. The errors listed are random and systematic, in this
order

\noindent $^e$Systematic change in \h0 resulting from (in order)
corrections for the local flow field, a dependence of the distance
moduli on Cepheid metallicity as in Kennicutt \etal (1998), and
population incompleteness bias in the distant sample. 
\end{table}

\section{Future Directions}

There is still substantial room for improvement in the result given
above.  One common dimension they all share is the assumption of a
50$\pm$3 kpc distance to the LMC. A distribution of LMC distances
compiled from the literature (Mould \etal 2000), if indeed peaked at
50 kpc, is not symmetric: red clump distances are as low as 43 kpc
(e.g., Stanek \etal 1999), while Mira variables define the upper
envelope at 55 kpc (Reid 1998). Mould \etal investigate the
consequences of replacing the adopted probability distribution with
this empirical, skewed compilation. As a result \h0 would increase by
4.5\%, and the associated error would jump to 12\%. While this is an
extreme, and somewhat unorthodox revision, it does illustrate the
rather heavy repercussion of this one assumption. The debate on the
LMC distance promises to be as heated as the controversy on \h0
itself, and a resolution might have to await the launch of SIM in
2005. In the meantime, an update on the LMC PL relation is long
overdue: the current calibration is based on 32 Cepheids, less than
the Cepheid sample size of several KP galaxies! Progress is being made
(e.g., Moffett 1998, Barnes \etal 1999, Tanvir \& Boyle 1999, Udalski
\etal 1999); within the KP, Kim Sebo is leading an effort 
which has so far produced {\it BVRIH\/} light curves for over 200 LMC
Cepheids (Sebo \etal 2000). Finally, systematics in the calibration
of the PL relation must also be explored in more galaxies having
Cepheid-independent distance estimates: promising starts are the
study conducted by Maoz \etal (1999) in NGC 4258, and the DIRECT
project targeting M31 and M33 (e.g., Mochejska \etal 1999).

The metallicity dependence of the Cepheid PL relation, and the
uncertainties in the photometric calibration of the WFPC2, closely
follow the LMC distance in generating the largest uncertainty in \h0.
While the latter will soon be better constrained (Stetson \etal 2000,
Saha \etal 2000), not much agreement has yet been reached for the
former (e.g., Alibert \etal 1999, Caputo \etal 1999, Storm \etal 
1998, Sasselov \etal 1997, Kochanek 1997). If the  mild, and only
marginally significant, metallicity dependence found for the KP by
Kennicutt \etal (1998) is applied to the Cepheid distances of the
local calibrators, the value of \h0 would decrease by 4.5\%.  

Further progress is also to be expected in improving the calibrator
samples for some of the secondary distance indicators. Sandage and
collaborators are still actively hunting down SNIa hosts. Indeed in
the near future three more calibrators will be added to the current
sample of six. One is a member of Fornax, and  will help to better
constrain the distance to the cluster needed for the FP calibration.
Finally, during this conference we learned that a
lot of effort is being spent in developing new distant samples
(Germany, Willick, Courteau, Giovanelli, Colless, this volume). In
particular, the Mount Stromlo Abell Cluster SN Search (see also Reiss
\etal 1998) coupled with the ongoing Cal\'an-Tololo and Asiago
surveys, promises to double the current number of distant SNe, with
the result that not only the Hubble diagram, but also dependences of
the peak magnitude on second parameters can be further refined. The
$I$-band Tully-Fisher sample of Dale \etal (1999) will push the method
to 25000 \kms, twice as far as the sample currently used by the KP,
while Fundamental Plane parameters for 80 new clusters are expected
from the EFAR project.

%
%
%

\clearpage

\begin{figure} 
\hskip -0.5in
\plotone{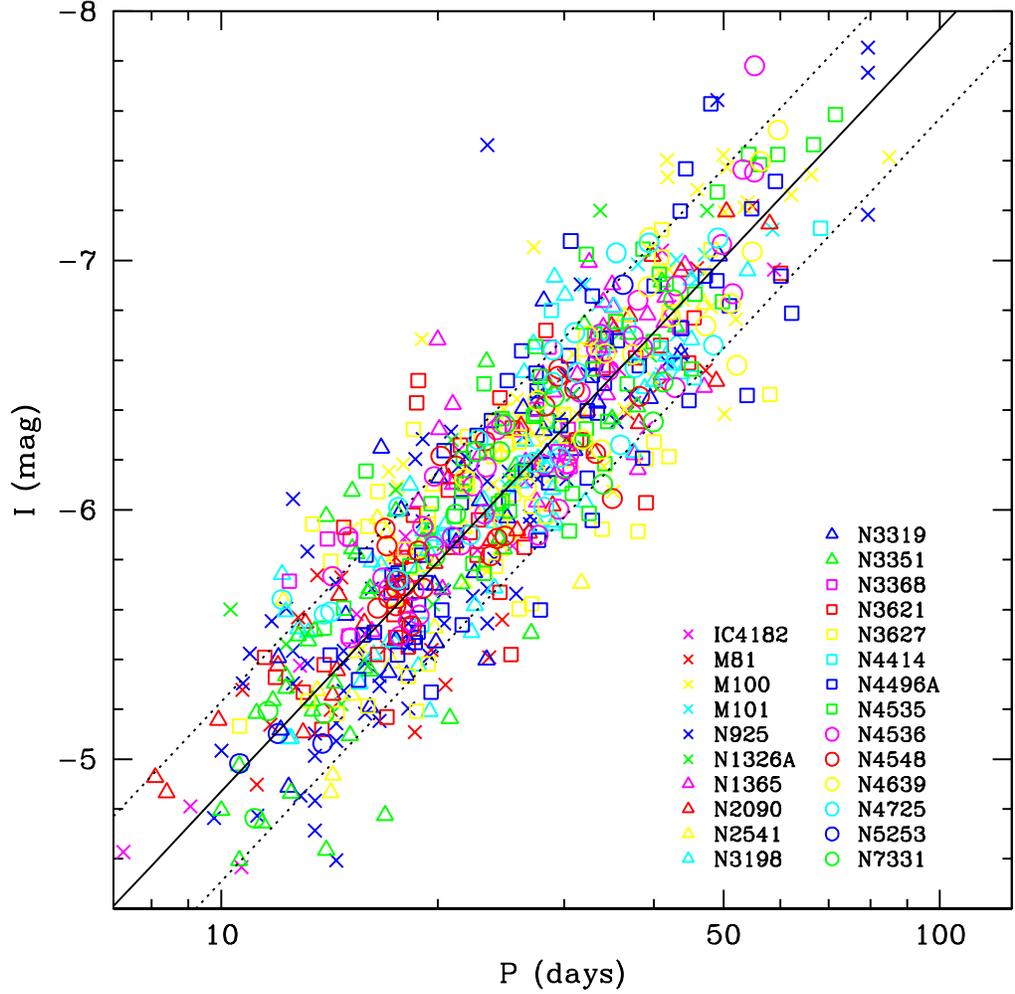} 
\caption{A composite $I$-band PL relation showing all Cepheids
discovered as part of the Key Project and other HST efforts
(references in Table 1). The count is just shy of 800
objects. Apparent magnitudes are converted to absolute magnitudes
using the distances in Table 1, and the reddening derived for each
galaxy. The solid line represents the calibrating LMC PL relation, and the
dotted lines its 2$\sigma$ uncertainty, deriving from the finite width of
the instability strip.} \label{fig-1}
\end{figure}

\begin{figure}
\hskip -0.5in
\plotone{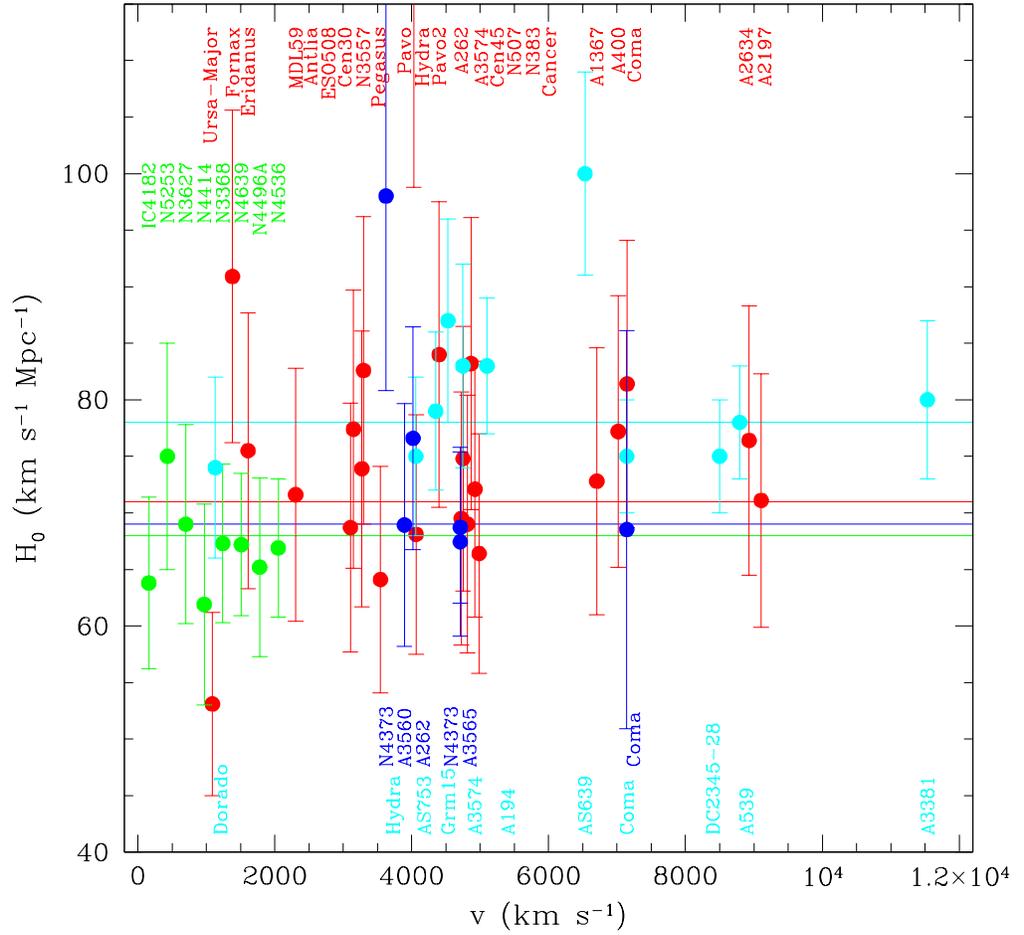} 
\caption{\h0 derived from the distant cluster sample for
Tully-Fisher (red), SBF (blue), and Fundamental Plane (cyan). For
SNIa (green) we plot the values of \h0 given by each of the local
calibrators when applied to the Hubble diagram produced by the distant
sample of Phillips \etal (1999). Velocities are CMB velocities for
Tully-Fisher, SBF and Fundamental Plane, heliocentric velocities for
the SNIa calibrators.} \label{fig-3}
\end{figure}

\end{document}